\documentstyle[12pt]{article}
\def\dspace{\baselineskip=0.3 in}
\begin{document}
\dspace

\centerline{\bf Tachyon as a Dark Energy Source}

\vspace{1cm}

\centerline{\bf S.K.Srivastava,}

\centerline{\bf Department of Mathematics,}

\centerline{\bf North Eastern Hill University,}

\centerline{\bf NEHU Campus,}

\centerline{\bf Shillong - 793022}

\centerline{\bf ( INDIA )}

\centerline{\bf e - mail : srivastava@nehu.ac.in; sushil@iucaa.ernet.in}

\vspace{2cm}

It is demonstrated that $dark$ $energy$, driven by tachyons, having
non-minimal coupling with curvature and self-interacting inverse cubic potential, decays
to $cold$ $dark$ $matter$ in the late accelerated universe. It is found that this
phenomenon yields a solution to ``cosmic coincidence problem''. 


\vspace{2cm}

The idea of tachyons is not new. It was proposed around 40 years back [1] and
some cosmological models also were developed [2]. Later on, these superluminal
particles were discarded for not being observed.  At the turn of the last century,
it has re-attracted attention of physicists appearing as condensates in some
of  string theories. After a series of papers by Sen[3], once again these particles were drawn into the arena of cosmology[4]. For not being observed, tachyons too are considered as good candidates for $dark$ $energy$ (DE),
apart from various DE models such as $quintessence$[5] and $k-essence$[6]
models, where these fields violate ``strong energy condition'' (SEC). Recently, some
other scalar field models for  DE have appeared where
``weak energy condition'' (WEC) is violated[7,8].

In the recent literature, study on tachyon scalar field $\phi$ has been done
with Born-Infeld lagrangian $ - V(\phi)\sqrt{ 1 -   g^{\mu\nu} \partial_{\mu} \phi \partial_{\nu} \phi }$ having minimal coupling with gravity. Later on, it was shown that this lagrangian can
also be treated as generalization of a relativistic particle lagrangian
[9]. Recently, another tachyon model, with minimal coupling, has been proposed with lagrangian $
W(\phi)\sqrt{    g^{\mu\nu} \partial_{\mu} \phi \partial_{\nu} \phi - 1 }$
(with $W(\phi)$ real). Using this lagrangian, it is argued that tachyon
scalars may be able to explore more physical situations than $quintessence$
[10].

For a non-tachyonic scalar $\psi$, having non-minimal coupling with curvature,
lagrangian is taken as $L^{\psi} = \sqrt{-g} [\frac{1}{2} g^{\mu\nu}
\partial_{\mu}\psi \partial_{\nu}\psi - \frac{1}{2} \xi R \psi^2 - V(\psi)]
(\xi $ being the coupling constant, $R$ and $V(\psi)$ being the curvature
scalar and potential respectively). In a similar way, it is reasonable to
explore dynamics of tachyon scalar $\phi$ also, taking its non-minimal coupling
with gravity, through the lagrangian

$$ L^{\phi} =  \sqrt{- g} \Big[ - V (\phi) \sqrt{ 1 -
  g^{\mu\nu} \partial_{\mu} \phi \partial_{\nu} \phi  + \xi R \phi^2} \Big] ,
  \eqno(1)$$
where $V(\phi)$ is the potential. 

 Non-minimal coupling of tachyon with gravity was also proposed by Piao $et$ $al$ [11] in a different manner, where a function of $\phi$ is coupled to Einstein-Hilbert lagrangian as

$$ S = \int {d^4x} \sqrt{- g} \Big[ \frac{f(\phi) R}{16 \pi G} + V (\phi) \sqrt{ 1 + \alpha^{\prime}
  g^{\mu\nu} \partial_{\mu}\phi \partial_{\nu}\phi } \Big]
  $$
Where $\alpha^{\prime}$ gives the string mass scale.

 Subject to the condition $
  1 - g^{\mu\nu} \partial_{\mu} \phi \partial_{\nu} \phi  >> \xi R \phi^2,$
  the lagrangian(1) looks like

$$ L_{\phi} \simeq  \sqrt{- g} \Big[  - V (\phi) \sqrt{ 1 -
  g^{\mu\nu} \partial_{\mu} \phi \partial_{\nu} \phi}  - \frac{1}{2} \frac{\xi
  V (\phi)\phi^2 R}{\sqrt{1 -
  g^{\mu\nu} \partial_{\mu} \phi \partial_{\nu} \phi}} , $$
which is similar to lagrangian of the action, taken by Piao et al with non-minimal coupling function
  $f(\phi)= - 8 \pi G \frac{\xi V (\phi)\phi^2 }{\sqrt{1 -
  g^{\mu\nu} \partial_{\mu} \phi \partial_{\nu} \phi}} $.

In what follows, investigations are made using the lagrangian (1) for the
tachyon scalar $\phi$.In the case of minimal coupling with gravity, tachyons behave as dust in the
late universe, as pressure for tachyon $p^{\phi} \to 0,$ when ${\dot \phi} =
\frac{d \phi}{dt} \to 1.$. In the model with non-minimal coupling, taken here,
$p^{\phi}$ does not vanish when ${\dot \phi} \to 1.$  Moreover, equation of
state parameter ${\rm w}^{\phi} = p^{\phi}/\rho^{\phi} < - 1/3,$   ( where
$\rho^{\phi}$ is the tachyon energy density) which causes accelerated
expansion of the universe predicted by recent experiments Ia supernova [12,
13] and WMAP (Wilkinson Microwave Anisotropy  Probe) [14 -16].  

It is found that DE density drecreases with expansion of the universe arousing
a question `` Where is  $dark$ $energy$ going ?''. Recently, in a
non-tachyonic case [17], it is proposed that there is no $dark$ $matter$ (DM)
in the beginning of the universe, but it is created due to decay of DE after
the universe starts expanding. As a result, DE density decreases and DM
density increases, in such a way that both become of the same  order at the
present age of the universe.   

Here, it is demonstrated that, in the late universe, $cold$ $dark$ $matter$
(CDM) is created due to decay of tachyon dark energy.  It is found that $ r(t)
= \rho^m(t)/ \rho^{\phi}(t)$   $(\rho^m (\rho^{\phi}) $ being CDM (DE)
density) grows with time, but keeping itself less than unity. Thus, it
provides a solution to ``cosmic coincidence problem'' CCP,  which asks how DE and
CDM densities are of the same order in the current universe [18].  This
mechanism is different from earlier ones [5], in which DE density is supposed to be
lower than the same for matter and radiation in the early universe, but become
comparable to the latter in the current universe. Moreover, it is also
different from the approach of Zimdahl $et$ $al.$ [19 - 21], where $r$
is shown stable in  late times.    Natural units (${\hbar} = c =1$) are used here.

The paper begins with Einstein's field equations

$$ R_{\mu\nu} - \frac{1}{2}g_{\mu\nu} R = - 8 \pi G (T_{\mu\nu}^{\phi} +
T_{\mu\nu}^{(m)} )  \eqno(2a)$$ 
with the gravitational constant $G = M_P^{-2} (M_P = 10^{19}$ GeV being the
  Planck mass) and  energy momentum tensor components for tachyons, given as

$$T_{\mu\nu}^{\phi} = (\rho^{\phi} + p^{\phi}) u_{\mu}u_{\nu} - p^{\phi}
g_{\mu\nu} \eqno(2b)$$
and
$$T_{\mu\nu}^{(m)} = (\rho^{(m)} + p^m) u_{\mu}u_{\nu} - p^mg_{\mu\nu},\eqno(2c)$$
where $\rho^{\phi}(\rho^m)$ and $p^{\phi}(p^m)$ are energy density and
pressure for tachyon(matter) respectively. Here, matter is CDM, so $p^m =
0$. $u^{\mu} = (1,0,0,0)$. For tachyonic perfect fluid $T^{(\phi) \mu}_{\mu} = (\rho^{\phi}, -
p^{\phi}, -  p^{\phi}, -  p^{\phi})$ are given by

\begin{eqnarray*}
 T_{\mu\nu}^{\phi} &=& - V (\phi) [ 1 -
   \bigtriangledown^{\rho} \phi \bigtriangledown_{\rho} \phi  + \xi R
   \phi^2]^{-1/2} \times \Big[ - \bigtriangledown_{\mu} \phi
   \bigtriangledown_{\nu} \phi + \xi R_{\mu \nu} \phi^2 \\ &&  + \xi (
   \bigtriangledown_{\mu} \bigtriangledown_{\nu} - g_{\mu\nu} {\Box} ) \phi^2
   - g_{\mu\nu} ( 1 -
   \bigtriangledown^{\rho} \phi \bigtriangledown_{\rho} \phi  + \xi R
   \phi^2) \Big]
\end{eqnarray*}
$$ \eqno(3)$$
derived from the lagrangian (1). Here $\bigtriangledown_{\mu}$ stands for
covariant derivative  $R_{\mu\nu}$ are Ricci tensor components .

Field equations for $\phi$ are obtained as 

$${\Box} \phi + \frac{2(\bigtriangledown^{\mu} \phi) (\bigtriangledown_{\rho}
  \phi) (\bigtriangledown^{\rho} \bigtriangledown_{\mu} \phi) - 2 \xi R \phi
  (\bigtriangledown^{\rho} \phi)( \bigtriangledown_{\rho} \phi) - \xi \phi^2
  g^{\mu\nu}(\bigtriangledown_{\mu}R) ( \bigtriangledown_{\nu} \phi)}{2(1 -
   \bigtriangledown^{\rho} \phi \bigtriangledown_{\rho} \phi  + \xi R
   \phi^2)} $$
$$ + \xi R \phi + \frac{V^{\prime}}{V} ( 1 + \xi R \phi^2 ) = 0, \eqno(4)$$ 
where $V^{\prime} (\phi) = \frac{d}{dx} V(\phi)$ and 
$${\Box} = \bigtriangledown^{\rho} \bigtriangledown_{\rho} = \frac{1}{\sqrt{-g}}\frac{\partial}{\partial x^{\mu}} ( \sqrt{-g} g^{\mu\nu} \frac{\partial}{\partial x^{\nu}}).$$

 Data of Ia Supernova [12, 13] and WMAP [14, 15, 16] indicate that we live in
 a spatially flat accelerated universe such that
${\ddot a}/ a > 0$ for the scale factor $a(t)$, given by the distance function

$$ dS^2 = dt^2 - a^2(t) [ dx^2 + dy^2 + dz^2 ] \eqno(5)$$
representing a homogeneous model of the universe. Hence, 

$$ \phi (x, t) = \phi (t) .\eqno(6)$$

The action $S$ shows that, in natural units, $\phi$ has mass dimension equal to $-1$ like time
$t$ . So, on the basis of dimensional considerations, it is
reasonable to take

$$  \phi (t) = A t .\eqno(7)$$
where $A$ is a dimensionless constant.

The potential is taken as

$$ V(\phi) = \sqrt{\frac{3 }{8 \pi G}} \phi^{-3},  \eqno(8)$$
as $G$ gives a natural scale in gravity.

\noindent \underline{\bf For the case $\xi \ne 0$}

Connecting eqs.(4) - (8), it is obtained that eq.(4) looks like

$$ 3 A H - \frac{ \xi A^3 t (2 R + t {\dot R}) }{2 [ 1 - A^2 + 2 \xi A^2 t^2 R]}
+ \xi A t R - \frac{3}{A t} ( 1 + \xi A^2 t^2 R ) = 0  , \eqno(9)$$
where $H = \frac{\dot a}{ a} $ with  ${\dot a } = \frac{da}{dt}.$

Eq.(9) admits the power-law solution

$$ a(t) = a_i ( t / t_i )^q  \eqno(10)$$
provided that

$$ 24 \xi q^2 A^2 - 3 A^2( 1 + 4 \xi ) q  + 3 = 0. \eqno(11)$$
In eq.(10), $t_i$ is the time, when decay of DE to CDM begins and $a_i$ is the corresponding scale
factor.

For the geometry, given by eq.(5), eqs.(3) yield energy density

$$\rho^{\phi} = T^{(\phi) 0}_0 = V(\phi) \frac{[ 1 + 6 \xi H \phi {\dot \phi} +3 \xi ({\dot H} + 3 H^2) \phi^2]
  }{\sqrt{1 - {\dot \phi}^2 + 6 \xi \phi^2 ({\dot H} + 2 H^2)}} \eqno(12a)$$
and isotropic pressure as

$$p^{\phi} =  -  V(\phi) \frac{[1 - {\dot \phi}^2 + \xi( 2 \phi {\ddot \phi} +
  2 {\dot \phi}^2 + 6 \xi H \phi {\dot \phi}) + \xi (5 {\dot H} + 9 H^2) \phi^2]  }{\sqrt{ 1 - {\dot \phi}^2 + 6 \xi \phi^2 ({\dot H} + 2 H^2)}} \eqno(12b)$$

Eqs.(12) show that, in minimal coupling case, $\rho^{\phi} = V(\phi)/\sqrt{1 -
  {\dot \phi}^2}$ and $p^{\phi} = -  V(\phi)\sqrt{1 -   {\dot \phi}^2}$ on
  taking $\xi = 0$ as obtained in refs.[3,4] for Sen's model. It is
  interesting to see that, on taking non-minimal coupling of $\phi$ with
  curvature  ($\xi \ne 0$) , $p^{(\phi)} \not\to 0$, when ${\dot \phi} \to 1$,
  contrary to Sen's model and Pio's way of taking non-minmal coupling, where $p^{(\phi)} \to 0$, when ${\dot \phi} \to 1$.

Due to dominance of DE over matter, eqs.(2) yield

$$ \frac{ R^1_1 - (1/2) R}{ R^0_0 - (1/2) R} = \frac{- p^{\phi}}{\rho^{\phi}}
= - {\rm w}^{\phi}. \eqno(13)$$

Connecting eqs.(7), (12) and (13), it is obtained that

$$ \frac{3 H^2 + 2 {\dot H}}{ 3 H^2} = \frac{ [1 - {\dot \phi}^2 + \xi( 2 \phi {\ddot \phi} +
  2 {\dot \phi}^2 + 6 \xi H \phi {\dot \phi}) + \xi (5 {\dot H} + 9 H^2)
  \phi^2] }{[ 1 + 6 \xi H \phi {\dot \phi} +3 \xi ({\dot H} + 3 H^2) \phi^2]}
  = - {\rm w}^{\phi}. \eqno(14)$$

Eqs.(7), (10) and (14) yield

$$ \frac{ 3 q - 2}{3 q} = \frac{1 - A^2  + \xi (2 + q + 9 q^2 ) A^2}{1 + 3 \xi
  A^2 (q + 2 q^2)} = - {\rm w}^{\phi}. \eqno(15)$$

Eliminating $A^2$ from eqs.(11) and (15), it is obtained that

$$ 12 \xi q^2 + 8 \xi q - 16 \xi + 1 = 0. \eqno(16)$$

Now subject to the condition $1 - {\dot \phi}^2 + 6 \xi ({\dot H} + 2 H^2) = 1
- A^2 + 6 \xi A^2 ( -q + 2 q^2 ) > 0$ ( to get $\rho^{\phi}$ and $p^{\phi}$
real), a set of solutions of eqs.(11) and (16) is obtained as

$$ q = \frac{5}{3}, \xi = - \frac{3}{92}, A = \sqrt{\frac{69}{125}} \quad {\rm
  and}\quad {\rm w}^{\phi} = - 0.6 . \eqno(17a,b,c,d)$$

As mentioned above, tachyon scalar field is a probable source of $dark$
$energy$. So, $\rho^{\phi}$ represents DE density. Eqs.(7),
(8), (10), (12) and (17) imply that
$$\rho^{\phi} = \sqrt{\Big(\frac{3 }{8 \pi G} \Big)} (A t)^{-3}
\frac{[1 + 3 \xi q (1 + 3 q) A^2]}{\sqrt{1 - A^2 + 6 \xi A^2 ( -q + 2 q^2 )}} 
 \eqno(18)$$

Eq.(18) shows that
DE density rolls down with growing time. Hence, it is reasonable to
propose that DE decays to CDM [17 ]. With $\rho^m$ as CDM density and $Q(t)$
as loss(gain) term for DE(CDM), this phenamenon is given by coupled equations

$${\dot \rho^{\phi}} + 3 H ( \rho^{\phi} + p^{\phi} ) = - Q(t) \eqno(19a)$$
and
$${\dot \rho^m} + 3 H  \rho^m =    Q(t) \eqno(19b)$$
derived from the Bianchi identities
$$ [T^{\mu\nu (\phi)} + T^{\mu\nu (m)}]_{;\nu} = 0.$$

From eq.(15),

$$ q = \frac{2}{3 ( 1 + {\rm w}^{\phi})}. \eqno(20)$$

Using the equation of state $p^{\phi} = {\rm w}^{\phi} \rho^{\phi}$ and
eq.(20) in eq.(19a), it is obtained that

$$ Q(t) =   \Big(\frac{1.84}{A^3 \sqrt{8 \pi G}} \Big) t^{- 4}
  \eqno(21)$$

As mentioned above, here, it is proposed that CDM is produced due to decay of
DE, so $\rho^{(m)}(t_i) = 0,$ where $t_i$ is the epoch when creation of CDM
begins.

Now eq.(19b) is integrated to 

$$\rho^m = \Big(\frac{1.84}{3 A^3 ( q - 1)\sqrt{8 \pi G}} \Big) t^{- 3} \Big[ 1
  - \Big(t_i/t \Big)^{3(q -1)} \Big]
  \eqno(22a)$$
using eqs.(10), (17),(21) and $\rho^m (t_i) = 0.$

Now,
$$ r(t) = \rho^m/ \rho^{\phi} = 0.5 \Big[ 1 - \Big(t_i/t \Big)^2 \Big] \eqno(22b)$$
for $q$ and $\rho^{\phi}$ from eqs.(17 a) and (18) respectively.
Current data of the universe $\rho^m_0 = 0.23 \rho_{cr,0}$ and $\rho^{\phi}_0 = 0.73 \rho_{cr,0}$ for
the current universe(where $\rho_{cr,0} = 3 H_0^2/{8 \pi G}, H_0 = h/t_0, t_0
= 13.7 {\rm Gyr}$ being the present age and parameter $ h = 0.72 \pm 0.08$) and
eq.(22b) yield

$$ t_i = 0.608 t_0  . \eqno(23)$$

Thus eqs.(22b) and (23) imply

$$ r(t) = 0.5 \Big[ 1 - 0.37 \Big(t_0/t \Big)^2 \Big] \eqno(24)$$
showing that $0 < r(t) < 1$ for $0.608 t_0 \le t.$ This result provides a
possible solution to CCP in the accelerated universe.

\noindent \underline{\bf For the case $\xi = 0$}

In this case, eqs.(12) and (13) yield

$$\frac{2 {\dot H}}{3 H^2 } = - {\dot \phi}^2 \eqno(25)$$
and eq.(4) looks like
$$ \frac{\ddot \phi}{1 - {\dot \phi}^2}  + 3 H {\dot \phi} - \frac{3}{\phi} =
0 \eqno(26)$$
for $V(\phi)$ given by eq.(8) [3,4].

The condition, for real $\rho^{\phi}$ and $p^{\phi}$, reduces to 
$$ {\dot \phi}^2 < 1 . \eqno(27)$$

Now, for slow roll-over of $\phi, {\ddot \phi} \ll 3 H {\dot \phi}$, so
eq.(26) is re-written as

$$  H {\dot \phi} = \phi^{-1}. \eqno(28)$$

Eqs.(25) and (28) yield

$$ H = \frac{4}{3} \frac{t_i^{3/2}}{\phi_i^2} t^{-1/2} \eqno(29a)$$
and
$$\phi = \phi_i \Big(t/t_i \Big)^{3/4}.  \eqno(29b)$$
Eq.(29a) yields $a(t) = a_0 exp\{\frac{8}{3} \frac{t_i^{3/2}}{\phi_i^2}
  t^{1/2}\}$ implying accelerated expansion.
Using these solutions in eq.(19a), it is obtained that

$$ Q(t) = \Big(\frac{3}{\phi} - \frac{9}{4 t} \Big) V(\phi)  \eqno(30)$$
with $\rho^{\phi} \simeq V(\phi)$.

Integration of eq.(19b) with $Q(t)$, given by eq.(30), yields

$$ \rho^m \approx \frac{3}{4} \sqrt{\frac{3}{8 \pi G}} \phi_i^{-2} t_i^{3/2}
t^{-5/2}. \eqno(31)$$

Eqs.(8) and (29b) yield
$$\rho^{\phi} \simeq V(\phi) = \sqrt{\frac{3}{8 \pi G}} \phi_i^{-3} \Big(t/t_i
\Big)^{-9/4} . \eqno(32)$$

Using the current data for $\rho^{\phi}$, given above, in eq.(32), it is
obtained that

$$ \phi^{-1} = \Big( 0.73 h^2 \sqrt{\frac{3}{8 \pi G}} \Big)^{1/3} t_0^{1/12}
t_i^{-3/4}.  \eqno(33)$$

Eqs.(31) - (33) yield

$$ \Big(\rho^m/\rho^{\phi} \Big)_{t = t_0} \approx \frac{3}{4} \Big(0.73 h^2
\sqrt{\frac{3}{8 \pi G}} \Big)^{-1/3} t_0^{-1/3} \approx 10^{-20}  \eqno(34)$$

This result ( for minimal coupling case) contradicts observational data for the current universe giving $
\Big(\rho^m/\rho^{\phi} \Big)_{t = t_0} \simeq 0.37.$

Thus, it is obtained that if tachyon dark energy decays to cold dark matter,
`` coincidence problem'' in the accelerated universe can be
 solved when tachyon scalar has non-minmal coupling to gravity, given by $\xi
 \ne 0$. This result is obtained for the universe using inverse cubic potential for $\phi$
 (tachyon scalar) and without dissipative term for
 matter. It is interesting to see that $\rho^m$ gradually increases and
 $\rho^{\phi}$ decreases in such a way that both are almost of the same order in the late uiniverse.

But in the case of minimal coupling ($\xi  = 0$), it is found that enough
amount of CDM is not created through decay of DE to get $\rho^m$ and
$\rho^{\phi}$ comparable. Here also, no dissipative term for matter is taken. So, for this case `` coincidence
 problem'' is not solved in the speeded-up universe. This is parrallel to
 results of refs.[20,21],where it is shown that, for the minimal case, CCP can
 not be solved for the accelerated universe without taking dissipative effects
 of CDM.

\centerline{\bf References}
\smallskip

\noindent [1] O.M.P.Bilaniuk, V.K.Deshpande and E.C.G.Sudarshan, Am. J. Phys. {\bf 30}, 718 (1962); O.M.P.Bilaniuk, V.K.Deshpande and E.C.G.Sudarshan, Physics Today {\bf 22}, 43 (1969); O.M.P.Bilaniuk and E.C.G.Sudarshan, Nature {\bf 223}, 386 (1969); G. Feinberg, Phys. Rev. {\bf 98}, 1089 (1967); E. Recami and R. Mignani, Riv. Nuovo Cimento, {\bf 4}, 209 (1974); A.F.Antippa and A. E. Everett, Phys. Rev. D {\bf 4}, 2098(1971); A.F.Antippa, Phys. Rev. D{\bf 11}, 724(1975); A. E. Everett, Phys. Rev. D, {\bf 13}, 785(1976); $idid$ {\bf 13}, 795(1976); K. H. Mariwalla, Am. J. Phys., {\bf 37}, 1281 (1969); L. Parker, Pys. Rev. {\bf 188}, 2287 (1969).
\bigskip

\noindent [2] J. C. Foster and J. R. Ray, J. Math. Phys. {\bf 13}, 979 (1972); J. R. Ray, Lett. Nuovo Cimento {\bf 12}, 249 (1975); S. K. Srivastava and M.P.Pathak, J. Math. Phys. {\bf 18}, 483 (1977); $ibid$ {\bf 18}, 2092 (1977); $ibid$ {\bf 19}, 2000 (1978); $ibid$ {\bf 23}, 1981 (1982); $ibid$ {\bf 24}, 966 (1983); $ibid$ {\bf 24}, 1311 (1983); $ibid$ {\bf 24}, 1317 (1983); $ibid$ {\bf 25}, 693 (1984).
\bigskip

\noindent [3] A. Sen, J. High Energy Phys. {\bf 04}, 048 (2002); {\bf 07}, 065 (2002); Mod. Phys. Lett. A {\bf 17}, 1799 (2002) and references therein.

\bigskip

\noindent [4] G. W. Gibbons, Phys. Lett. B {\bf 537}, 1 (2002);
T. Padmanabhan, Phys. Rev. D {\bf 66}, 021301 (2002); T. Padmanabhan and
T. R. Choudhury, Phys. Rev. D {\bf 66}, 081301 (2002); M. Fairbairn and
M. H. G. Tytgat, Phys. Lett. B {\bf 546}, 1 (2002); A. Feinstein, D {\bf 66},
063511 (2002); D. Choudhury, D. Ghosal, D.P.Jatkar and S. Panda, Phys. Lett. B
{\bf 544}, 231 (2002); A. Frolov, L. Kofman and A. A. Starobinsky,
Phys. Lett. B {\bf 545}, 8 (2002); L. Kofman and A. Linde, J. High Energy
Phys. {\bf 07}, 065 (2002); G. Shiu and I. Wasserman,  Phys. Lett. B {\bf
  541}, 6 (2002); L. R. W. Abramo and F. Finelli, Phys. Lett. B, {\bf 575},
165 (2003); M. Sami, P. Chingangbam and T. Qureshi, Phys. Rev. D {\bf 66}, 043530 (2002).

\bigskip

\noindent [5] B. Ratra and P. J. E.Peebles, Phys.Rev. {\bf D 37}, 3406(1988); C. Wetterich, Nucl. Phys. {\bf B 302}, 668 (1988); J. Frieman, C.T. Hill, A. Stebbins and I. Waga, Phys. Rev. Lett. {\bf 75}, 2077 (1995); P. G. Ferreira and M. Joyce, Phys.Rev. {\bf D 58}, 023503(1998); I. Zlatev, L. Wang and P. J. Steinhardt, Phys. Rev. Lett. {\bf 82}, 896 (1999); P. Brax and J. Martin,  Phys.Rev. {\bf D 61}, 103502(2000);
 L. A. Ure${\tilde n}$a-L${\acute o}$pez and T. Matos Phys.Rev. {\bf D 62}, 081302(R) (2000); T. Barrriro, E. J. Copeland and N.J. Nunes, Phys.Rev. {\bf D 61}, 127301 (2000); A. Albrecht and C. Skordis,  Phys. Rev. Lett. {\bf D 84}, 2076 (2000); V. B. Johri,  Phys.Rev. {\bf D 63}, 103504 (2001); J. P. Kneller and L. E. Strigari, astro-ph/0302167; F. Rossati, hep-ph/0302159; V. Sahni, M. Sami and T. Souradeep,  Phys. Rev. {\bf D 65}, 023518 (2002); M. Sami, N. Dadhich and Tetsuya Shiromizu, hep-th/0304187.

\bigskip

\noindent [6] C. Armendariz-Picon, T. Damour and V. Mukhanov, Phys. Lett. {\bf B 458}, 209 (1999); T. Chiba, T. Okabe and M. Yamaguchi, Phys. Rev. {\bf D 62}, 023511 (2000)

\bigskip

\noindent [7] V.Faraoni, Phys. Rev.D{\bf 68} (2003) 063508; R.A. Daly $et$
$al.$, astro-ph/0203113; R.A. Daly and E.J. Guerra, Astron. J. {\bf 124}
(2002) 1831;  R.A. Daly, astro-ph/0212107; S. Hannestad and E. Mortsell,
Phys. Rev.D{\bf 66} (2002) 063508; A. Melchiorri $et$ $al.$, Phys. Rev.D{\bf
  68} (2003) 043509; P. Schuecker $et$ $al.$, astro-ph/0211480;
R.R. Caldwell, Phys. Lett. B {\bf 545} (2002) 23; R.R. Caldwell,
M. Kamionkowski and N.N. Weinberg, Phys. Rev. Lett. {\bf 91} (2003) 071301;
H. Ziaeepour, astro-ph/0002400 ; astro-ph/0301640; P.H. Frampton and
T. Takahashi,  Phys. Lett. B {\bf 557} (2003) 135;  P.H. Frampton,
hep-th/0302007;  S.M. Carroll $et$ $al.$, Phys. Rev.D{\bf 68} (2003) 023509;
J.M.Cline $et$ $al.$, hep-ph/0311312;  U. Alam , $et$ $al.$, astro-ph/0311364
; astro-ph/0403687; O. Bertolami, $et$ $al.$, astro-ph/0402387;  P. Singh,
M. Sami $\&$ N. Dadhich, Phys. Rev.D{\bf 68} (2003) 023522; M. Sami $\&$
A. Toporesky, gr-qc/0312009;  B. McInnes, JHEP, {\bf 08} (2002) 029;
hep-th/01120066;  V. Sahni $\&$ Yu.V.Shtanov, JCAP {\bf 0311} (2003) 014;
astro-ph/0202346;  Pedro F. Gonz$\acute a$lez-D$\acute i$az, Phys. Rev.D{\bf
  68} (2003) 021303(R); J. D. Barrow, Class. Quan. Grav. {\bf 21}, L 79 - L82
(2004) [gr-qc/0403084];  S. Nojiri and S. D. Odintsov, hep-th/0303117; hep-th/0306212; E. Elizalde, S. Noriji and S. D. Odintsov,  Phys. Rev.D{\bf 70}, 043539 (2004); hep-th/0405034, S. Nojiri and S. D. Odintsov, Phys. Lett. B {\bf 595}, 1 (2004); hep-th/0405078; hep-th/0408170  .

\bigskip

\noindent [8] S. K. Srivastava, astro-ph/0407048; V. K. Onemli and R. P. Woodard, Class. Quant. Grav. {\bf 19}, 4607 (2002) [gr-qc/0204065]; gr-qc/0406098; T. Brunier, V. K. Onemli and R. P. Woodard, gr-qc/0408080.

\bigskip

\noindent [9] J. S. Bagla, H. K. Jassal and T. Padmanabhan, Phys. Rev. D {\bf 67}, 063504 (2003).

\bigskip

\noindent [10] V. Gorini, A. Kamenshchik, U. Moschella and V. Pasquir,
Phys. Rev. D {\bf 69}, 123512 (2004); hep-th/0311111.

\bigskip

\noindent [11] Yun-Song Pio $et$ $al$, Phys. Lett. {\bf B 570},1 (2003); hep-ph/0212219.

\bigskip

\noindent [12] S. Perlmutter $et$ $al$., Astrophys. J., {\bf 517}, 565 (1999).

\bigskip

\noindent [13] A.G. Riess $et$ $al$., Astron. J., {\bf 116}, 1009 (1998).

\bigskip

\noindent [14] D. N. Spergel $et$ $al$., astro-ph/0302209.

\bigskip

\noindent [15] L. Page $et$ $al$., astro-ph/0302220.

\bigskip

\noindent [16] A. B. Lahnas, N. E. Nanopoulos and D. V. Nanopoulos, Int. Jour. Mod. Phys. D, {\bf 12}, 1529 (2003).

\bigskip

\noindent [17] S. K. Srivastava, hep-th/0404170.

\bigskip

\noindent [18] P. J. Steihardt, Cosmological challenges for the 21st century, in : V. L. Fitch, D.R.Marlow (edts.), Critical Problems in Physics, Princeton Univ. Press, Princeton, NJ, 1997. 

\bigskip

\noindent [19] W. Zimdahl, D. Pav${\acute o}$n and L. P. Chimento, Phys. Lett. B {\bf 521}, 133 (2001).

\bigskip

\noindent [20] L. P. Chimento, A. S. Jakubi, D. Pav${\acute o}$n and W. Zimdahl, Phys. Rev. D {\bf 67}, 083513 (2003); $ibid$ {\bf 67}, 087302 (2003) .

\bigskip

\noindent [21] R. Herrera, D. Pav${\acute o}$n and W. Zimdahl, astr0-ph/0404086. 

\end{document}